\documentclass[a4paper,fleqn]{cas-dc}

\usepackage[numbers]{natbib}
\usepackage{widetext}

\def\tsc#1{\csdef{#1}{\textsc{\lowercase{#1}}\xspace}}
\tsc{WGM}
\tsc{QE}
\tsc{EP}
\tsc{PMS}
\tsc{BEC}
\tsc{DE}
\begin{document}
\let\WriteBookmarks\relax
\def\floatpagepagefraction{1}
\def\textpagefraction{.001}
\shorttitle{}
\shortauthors{}
%\begin{frontmatter}

\title [mode = title]{Effect of interfacial spin mixing conductance on gyromagnetic ratio of Gd substituted Y\textsubscript{3}Fe\textsubscript{5}O\textsubscript{12}} 
\tnotemark[1]

\tnotetext[1]{This document is the results of the research project funded by Ministry of Research and Technology of the Republic of Indonesia through PDUPT Grant No. NKB-175/UN2.RST/HKP.05.00/2021.}

\author[1]{Adam B. Cahaya}[
orcid=0000-0002-2068-9613]
\cormark[1]
\ead{adam@sci.ui.ac.id}

\credit{Conceptualization, Methodology, Writing - Review & Editing}

\author[2]{Anugrah Azhar}[
orcid=0000-0003-1068-1498]
\credit{Data curation}

\author[1]{Dede Djuhana}[
orcid=0000-0002-2025-0782]
%\ead{aziz.majidi@sci.ui.ac.id}

\credit{Funding acquisition }

\author[1]{Muhammad Aziz Majidi}[
orcid=0000-0002-0613-1293]
%\ead{aziz.majidi@sci.ui.ac.id}

\credit{Supervision, Funding acquisition }
\address[1]{Department of Physics, Faculty of Mathematics and Natural Sciences, Universitas Indonesia, Depok 16424, Indonesia}

\address[2]{Physics Study Program, Faculty of Sciences and Technology, Syarif Hidayatullah State Islamic University Jakarta, Tangerang Selatan 15412, Indonesia}

\begin{abstract}
Due to its low intrinsic damping, Y$_3$Fe$_5$O$_{12}$ and its substituted variations are often used for ferromagnetic layer at spin pumping experiment. Spin pumping is an interfacial spin current generation in the interface of ferromagnet and non-magnetic metal, governed by spin mixing conductance parameter $G^{\uparrow\downarrow}$. $G^{\uparrow\downarrow}$ has been shown to enhance the damping of the ferromagnetic layer. The theory suggested that the effect of $G^{\uparrow\downarrow}$ on gyromagnetic ratio only comes from its negligible imaginary part. In this article,  we show that the different damping of ferrimagnetic lattices induced by $G^{\uparrow\downarrow}$ can affect the gyromagnetic ratio of Gd-substituted Y$_3$Fe$_5$O$_{12}$.
\end{abstract}

\maketitle

\begin{small}
\section{Introduction}
One of the focuses of spintronics, the research area about the manipulation of spin degree of freedom, is the manipulation of magnetic moment by spin current and vice-versa \cite{PhysRevB.72.024426,HIROHATA2020166711}.
At magnetic interface, a spin current can be generated from a ferromagnetic layer to non-magnetic metallic layer by spin pumping \cite{PhysRevB.66.224403}. 
The pumped spin current arises from the exchange interaction between the spin of ferromagnetic layer and the conduction spin of the non-magnetic metal \cite{PhysRevB.96.144434}. The polarization of the pumped spin current depends on the dynamics of magnetic moments at the ferromagnetic layer  \cite{PhysRevLett.88.117601}
\begin{align}
\textbf{J}=\mathrm{Re} G^{\uparrow\downarrow} \textbf{m}\times \dot{\textbf{m}}- \mathrm{Im}G^{\uparrow\downarrow} \dot{\textbf{m}},
\end{align}
where $\textbf{m}$ is the normalized magnetization direction and $G^{\uparrow\downarrow}$ is the interfacial spin mixing conductance \cite{PhysRevLett.111.176601}. While $G^{\uparrow\downarrow}$ generally has a complex value, its imaginary part is significantly smaller \cite{PhysRevB.76.104409,Sasage2010}.

Due to its low intrinsic damping \cite{6882836,Serga_2010}, Y$_3$Fe$_5$O$_{12}$ (YIG) \cite{PhysRevLett.107.066604,PhysRevB.99.220402,Burrowes2012} and its substituted variations\cite{PhysRevB.87.104412,PhysRevMaterials.2.094405} are often used for ferromagnetic layer at spin pumping experiment. 
It has been observed that the spin mixing conductance can be enhanced by substituting Y in ferrimagnetic Y$_3$Fe$_5$O$_{12}$  with rare earths such as Gd \cite{Geprags2016,Iwasaki2019,ortiz2021ultrafast}. 
Furthermore, the polarization switch of the spin current near the magnetization compensation point of Gd$_3$Fe$_5$O$_{12}$ is often studied \cite{Geprgs2016,PhysRevB.99.024417}. The magnetization compensation points occur because Gd$^{3+}$ and Fe$^{3+}$ ions in ferrimagnetic Gd$_3$Fe$_5$O$_{12}$ are antiferromagnetically coupled.

Beside using ferromagnetic resonance, spin pumping can be excited using temperature gradient $\Delta T$ and produce spin Seebeck voltage \cite{PhysRevB.81.214418,7111309}
\begin{align}
V\propto \frac{\gamma \mathrm{Re} G^{\uparrow\downarrow}}{M_s}\Delta T, \label{Eq.SSE}
\end{align}
where $\gamma$ and $M_s$ are the gyromagnetic ratio and saturation magnetization of ferromagnetic layer, respectively. The proportionality constant only depends on the properties of the non-magnetic metal. Ref.~\cite{PhysRevB.87.104412} shows that the magnitude of $G^{\uparrow\downarrow}$ at the interface of Gd-substituted Y$_3$ Fe$_5$O$_{12}|$Pt is dominantly originated from the magnetization of Fe $G^{\uparrow\downarrow}\propto M_{\rm Fe}$. However, the linear relation of $\gamma$ and $V$ was not confirmed.

Reciprocally, spin mixing at the interface gives a torque on the magnetization in the form of spin transfer torque { due to spin accumulation $\boldsymbol{\mu}_s$ \cite{PhysRevB.77.224419}. 
\begin{align}
\tau=\mathrm{Re}G^{\uparrow\downarrow} \textbf{m}\times\left(\textbf{m}\times\boldsymbol{\mu}_s\right) - \mathrm{Im}G^{\uparrow\downarrow} \textbf{m}\times\boldsymbol{\mu}_s \label{Eq.stt}
\end{align}
}
Spin transfer torque can be used for manipulation of the magnetization of the ferromagnetic layer \cite{Stiles,PhysRevLett.99.066603}. The real part of spin mixing conductance  $\mathrm{Re}G^{\uparrow\downarrow}$ has been shown to increase the Gilbert damping of the magnetization \cite{PhysRevLett.88.117601}. 
\begin{align}
\gamma\alpha=\gamma\alpha^{(0)} + M_j \mathrm{Re}G^{\uparrow\downarrow}, 
\end{align}
On the other hand, $\mathrm{Im}G^{\uparrow\downarrow}$ has been predicted to reduce the gyromagnetic ratio \cite{PhysRevB.96.144434,PhysRevLett.88.117601}. 
\begin{align}
\frac{1}{\gamma}=\frac{1}{\gamma^{(0)}} + M_j \mathrm{Im}G^{\uparrow\downarrow}, 
\end{align}
However, the effect of $\mathrm{Re}G^{\uparrow\downarrow}$ to the gyromagnetic ratio is not well-studied.

In this article, we aim to study the effect of $G^{\uparrow\downarrow}$ on the gyromagnetic ratio of Gd substituted Y$_{3}$Fe$_5$O$_{12}$, assuming negligible $\mathrm{Im}G^{\uparrow\downarrow}\to 0$. While $G^{\uparrow\downarrow}$ has been predicted to only increase the damping, it can also affect the gyromagnetic ratio, because the effective gyromagnetic ratio of a ferrimagnet is determined on the damping parameters of each magnetic lattice \cite{Wangness}. By studying the damping increase due to $G^{\uparrow\downarrow}$ of the interface in Sec.~\ref{Sec:interface} and the coupled dynamics of two magnetic lattices in Sec.~\ref{Sec:LL}, we can describe the effect of spin mixing conductance on the effective gyromagnetic ratio of Gd substituted YIG in Sec.~\ref{Sec:YGdIG} and show that $\gamma$ in Eq.~\ref{Eq.SSE} should be the $G^{\uparrow\downarrow}$-corrected gyromagnetic ratio.

\newpage
\section{Methods}
\subsection{Damping torque due to interfacial spin mixing }
\label{Sec:interface}

In second quantization, the interactions of conduction spin of non-magnetic metal near the interface with $n$-th spin $\textbf{S}_n$ of ferromagnet layer can be written with the following  $s-d$ Hamiltonian \cite{PhysRevB.103.094420}
\begin{align}
\mathcal{H}=&\sum_{\textbf{p}\alpha} \epsilon_\textbf{p}a_{\textbf{p}\alpha}^\dagger a_{\textbf{p}\alpha}
-\gamma_e\sum_{\textbf{p}\alpha\beta} \textbf{H}\cdot\boldsymbol{\sigma}_{\alpha\beta} a_{\textbf{p}\alpha}^\dagger a_{\textbf{p}\beta}\notag\\
&- J \sum_{n\textbf{pq}\alpha\beta} \textbf{S}_n \cdot\boldsymbol{\sigma}_{\alpha\beta} a_{\textbf{p}+\textbf{q}\alpha}^\dagger a_{\textbf{p}\beta},\label{Eq.Hamiltonian}
\end{align}
where $\gamma_e$ is the gyromagnetic ratio of free electron, $a_{\textbf{p}\alpha}^\dagger (a_{\textbf{p}\alpha})$ is the creation (annihilation) operator of conduction electron with wave vector $\textbf{p}$ and spin $\alpha$,  $\boldsymbol{\sigma}$ is Pauli vectors, $\epsilon_\textbf{p}=\hbar^2p^2/2m$ is the energy of conduction electron and $J$ is the exchange constant.

In linear response regime, the exchange interaction dictates that the spin density of the conduction electron responds linearly to perturbation due to exchange interaction\cite{PhysRevB.96.144434,Cahaya2021}
\begin{align}
{\sigma}_i(\textbf{r})=& \sum_{\textbf{pq}\alpha\beta} e^{i\textbf{q}\cdot \textbf{r}}\boldsymbol{\sigma}_{\alpha\beta} a_{\textbf{p}+\textbf{q}\alpha}^\dagger a_{\textbf{p}\beta} \notag\\
=& J\sum_n \int d\textbf{r} dt \chi_{ij} (\textbf{r}-\textbf{r}',t-t') S_{nj}(\textbf{r}',t'),
\end{align}
where $i,j\in\left\{x,y,z\right\}$. The susceptibility
\begin{align}
\chi_{ij} (\textbf{r},t)=& \frac{i}{\hbar}\Theta (t) \left< \left[\sigma_i(\textbf{r},t),\sigma_j(\textbf{0},0)\right]\right>
\end{align}
can be determined by evaluating its time derivation
\begin{align}
\frac{\partial\chi_{ij}(\textbf{r},t)}{\partial t}=\frac{i}{\hbar}\Theta (t) \left< \left[\frac{1}{i\hbar}\left[\sigma_i(\textbf{r},t),\mathcal{H}_0\right],	\sigma_j(\textbf{0},0)\right]\right>.
\end{align}
By setting the first two terms in Hamiltonian in Eq.~\ref{Eq.Hamiltonian} as the unperturbed $\mathcal{H}_0$, the susceptibility can be evaluated
$
\chi_{ij}(\textbf{r},t)= \sum_{\textbf{pq}} e^{i\textbf{q}\cdot \textbf{r}-i\omega t}\chi_{ij}
(\textbf{p},\textbf{q},\omega),
$
we can derive the exact expression of $\chi_{ij}$ in the static limit $\omega\to0$ for all $i,j$ combination

\begin{align}
&\sum_\textbf{p}\left(
\begin{array}{ccc}
\chi_{xx}(\textbf{p},\textbf{q},0) & \chi_{xy}(\textbf{p},\textbf{q},0) & \chi_{xz}(\textbf{p},\textbf{q},0) \\
\chi_{yx}(\textbf{p},\textbf{q},0) & \chi_{yy}(\textbf{p},\textbf{q},0) & \chi_{yz}(\textbf{p},\textbf{q},0) \\
\chi_{zx}(\textbf{p},\textbf{q},0) & \chi_{zy}(\textbf{p},\textbf{q},0) & \chi_{zz}(\textbf{p},\textbf{q},0)
\end{array}
\right)\notag\\
&= 
\left(
\begin{array}{ccc}
\chi(q) & \gamma_e H_z \varphi(q) & \gamma_eH_y \varphi(q) \\
-\gamma_eH_z \varphi(q) & \chi(q) & \gamma_eH_x \varphi(q)\\
\gamma_eH_y \varphi(q) & -\gamma_eH_x \varphi(q) & \chi(q)
\end{array}
\right)
\end{align}
such that
\begin{align}
\chi_{ij}(\textbf{r},t)=&\sum_\textbf{q} e^{i\textbf{q}\cdot \textbf{r}-i\omega t}\sum_{\textbf{p}}\chi_{ij}(\textbf{p},\textbf{q},0) \notag\\
=& \delta (t)\sum_\textbf{q} e^{i\textbf{q}\cdot \textbf{r}}\Big(\delta_{ij}\chi(q) +\epsilon_{ijk}\gamma_eH_k \varphi(q)\Big).
\end{align}
One can see that the susceptibility is anisotropic \cite{Cahaya_2021}. In the limit of small magnetic field $H\ll \epsilon_F$, the induced spin density takes the following form 
\begin{align}
\boldsymbol{\sigma}(\textbf{r})=\sum_{n\textbf{k}} e^{i\textbf{k}\cdot \textbf{r}}J\left(\chi(k) \textbf{S}_n  + \gamma_e \varphi(k) \textbf{S}_n\times \boldsymbol{H} \right), \label{Eq.20}
\end{align}
where $\chi(k)$ is the static susceptibility of a metal  
\begin{align}
\chi
=& \mathcal{N}(\epsilon_F) \left(1+\frac{4k_F^2-q^2}{4k_Fk}\log\left|\frac{k+2k_F}{k-2k_F}\right|\right)
\end{align}
and 
\begin{align}
\varphi(k)=&\lim_{\eta\to 0}\sum_{\textbf{p}}\frac{f_{\textbf{p}}-f_{\textbf{p}+\textbf{k}}}{\left(\epsilon_{\textbf{p}+\textbf{k}}-\epsilon_{\textbf{p}}+i\eta\right)^2}\notag\\
=&\mathcal{N}^2(\epsilon_F)\frac{\pi^3}{k_F^2\hbar } \frac{\Theta(2k_F-k)}{k}. 
\end{align}
{ $\varphi$ is the anisotropic susceptibility that generates a term in $\boldsymbol{\sigma}$ that is non-collinear to $\textbf{S}_n$. Here $\mathcal{N}(\epsilon_F)$ is the density of state at Fermi level. $\varphi$ term generates a spin transfer torque on spin $\textbf{S}_n$ \cite{PhysRevB.103.094420} 

\begin{align}
\tau=& % \frac{1}{i\hbar}\left[\textbf{S}_n,\mathcal{H}\right]=
\sum_n\gamma_n J\textbf{S}_n\times\boldsymbol{\sigma(\textbf{0})}\notag\\
=&\left( J^2\sum_\textbf{k}\varphi(k)\right)\sum_n \gamma_n\textbf{S}_n\times \left(\textbf{S}_n\times\gamma_e\textbf{H}\right). \label{Eq.torque}
\end{align}
Since $\textbf{S}_n=S_n\textbf{m}$ and $\gamma_e \textbf{H}$ is a spin accumulation, by comparing Eqs.~\ref{Eq.torque} and \ref{Eq.stt} one can see that $\varphi$ is related to spin mixing conductance $G^{\uparrow\downarrow}=\sum_j G_j^{\uparrow\downarrow}$, where the spin mixing conductance for $j-$th lattice is
\begin{equation}
G_j^{\uparrow\downarrow}=N_jS_j^2J^2\sum_\textbf{k}\varphi(k)%= \frac{N_j\pi S_j^2J^2\mathcal{N}^2(\epsilon_F)}{\hbar}
,\label{Eq.GproptoS2}
\end{equation}
where $N_j$ is number of spin at the interface. This torque increase the damping torque on the total magnetic moment $\textbf{M}=M_sV\textbf{m}$ of the whole volume of the ferromagnetic layer
\begin{equation}
\frac{d\textbf{M}}{dt}=\sum_n \gamma_j G_j^{\uparrow\downarrow}\textbf{m}\times\left(\textbf{m}\times \textbf{H}\right)\label{Eq.15},
\end{equation}
can be written in a normalized form 
\begin{equation}
\frac{d\textbf{m}}{dt}= \frac{1}{M_s V} \sum_n \gamma_j G_j^{\uparrow\downarrow}\textbf{m}\times\left(\textbf{m}\times \textbf{H}\right) 
\end{equation}
where $M_s$ is magnetization saturation, $V=Ad$ is volume of the magnetic layer. One can see the damping due to spin mixing conductance is inversely proportional to thickness $d$. 
For YIG, Ref.~\cite{PhysRevB.96.144434} estimate the value per unit area to be $
G_{\rm YIG}^{\uparrow\downarrow}/A\sim \textrm{\AA}^{-2}$. When Y is substituted with Gd, the spin mixing conductance should include the contributions from all magnetic lattice \cite{CAHAYA2022169248}.

}
\subsection{Landau-Lifshitz equation of ferrimagnet} 
\label{Sec:LL}
The dynamics of magnetic moment of $j$-th magnetic lattice $\textbf{M}_j$ ($j=1,2$) in a ferrimagnet is governed by Landau-Lifshitz equation \cite{Wangness}.
\begin{align}
\frac{d\mathbf{M}_j}{dt}=&-\gamma_i\mathbf{M}_j\times\mathbf{H}_{j}%
-\frac{\alpha_j\gamma_j}{M_j}\mathbf{M}_j\times\left(\mathbf{M}_j\times\mathbf{H}_{j}\right), \label{Eq.LL}
\end{align}
where $\mathbf{H}_{j}$ is the effective magnetic field felt by $\textbf{M}_j$ and $\alpha_j$ is its dimensionless damping parameter. $\textbf{H}_j$ consists of external magnetic field $H_0$ and molecular field due to coupling with another magnetic lattice
\begin{align}
\textbf{H}_{j}=& \textbf{H}_0 + \lambda \textbf{M}_{k\neq j}.
\end{align}
$\lambda$ is coupling constant between magnetic lattices. The $\mathbf{M}_j\times\left(\mathbf{M}_j\times\mathbf{H}_{j}\right)$ term in Eq.~\ref{Eq.LL} is the damping torque \cite{Wolf_1961}, that include the contribution of spin mixing conductance in Eq.~\ref{Eq.15}
{
\begin{equation}
%\frac{\left(\alpha_j-\alpha_j^{(0)}\right)\gamma_j}{M_j}\textbf{M}_j\times\left(\textbf{M}_j\times\textbf{H}\right)&=&\frac{N_j\gamma_j G_j^{\uparrow\downarrow}}{M_j^2}\textbf{M}_j\times\left(\textbf{M}_j\times\textbf{H}\right)\\
\left(\alpha_j-\alpha_j^{(0)}\right)=\frac{N_jG_j^{\uparrow\downarrow}}{M_jV} , \label{Eq.alphaS}
\end{equation}
where $N_j$ is number of spin at the interface, $\alpha_j^{(0)}$ is the intrinsic damping of $i$-th magnetic lattice of the magnetic layer. One can see the damping enhancement is inversely proportional to thickness of the ferromagnetic layer. 
}

Here we note that the damping torque could take $\textbf{M}\times\dot{\textbf{M}}$ form as in the Landau-Lifshitz-Gilbert equation \cite{Lakshmanan2011}. However, Ref.~\cite{Kittel1959,Kittel1960} shows that Eq.~\ref{Eq.LL} has better agreement with the experiment data for rare earth garnet in large damping limit, which is appropriate for spin pumping setup that has large damping. 

\begin{figure}
\includegraphics[width=\columnwidth]{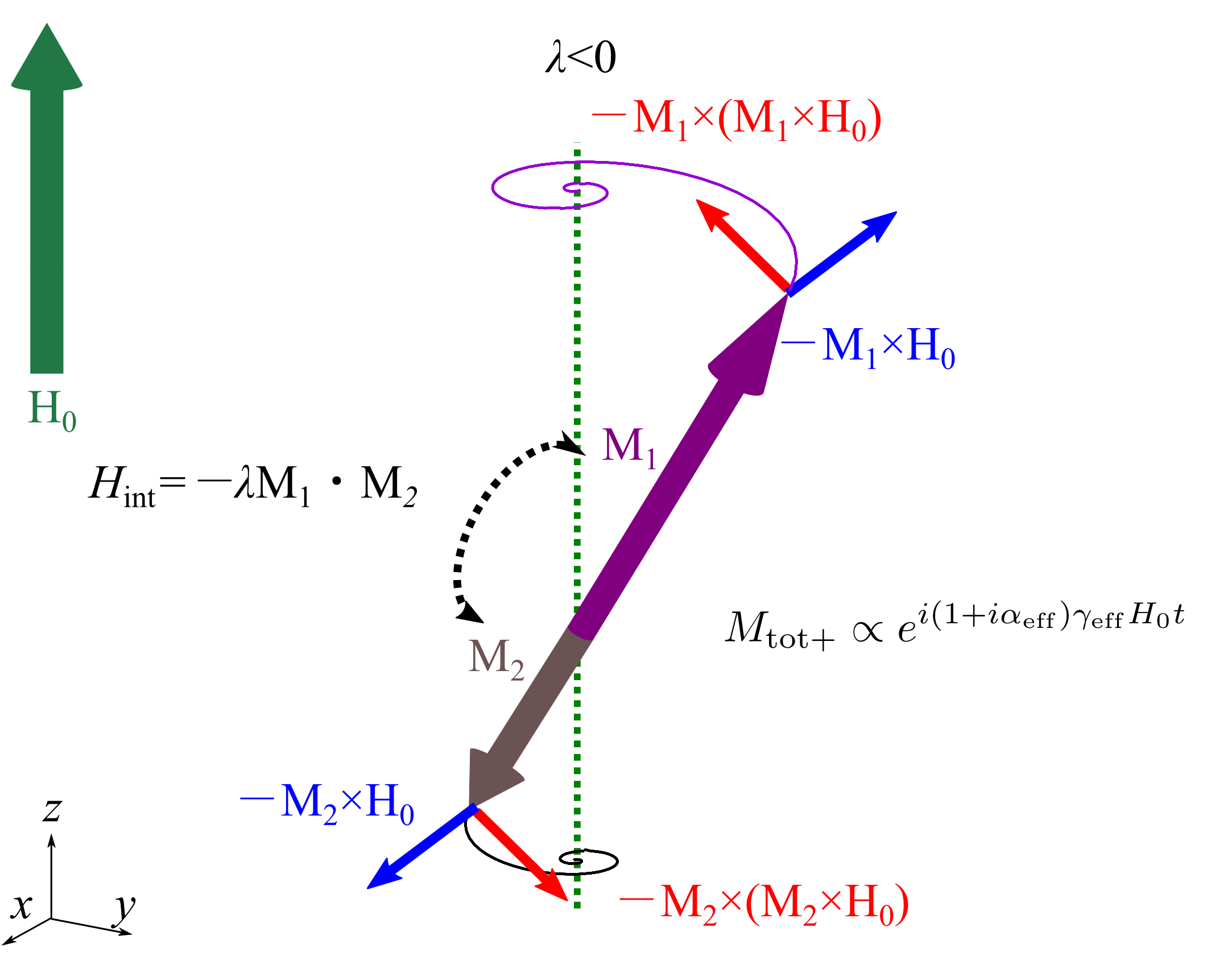}
\caption{Ferrimagnet with two magnetic lattices $\textbf{M}_1$ and $\textbf{M}_2$ under an external magnetic field $\textbf{H}_0$.  Due to magnetic interaction $H_\mathrm{int}=-\lambda \textbf{M}_1\cdot\textbf{M}_2$, the dynamics of $\textbf{M}_1$ and $\textbf{M}_2$ are coupled as a total magnetization with effective gyromagnetic ratio $\gamma_\mathrm{eff}$ and effective damping parameter $\alpha_\mathrm{eff}$. When $\lambda>0$ the coupling is ferromagnetic. On the other hand, when $\lambda<0$, the coupling is antiferromagnetic.
\label{Fig.Ferri}}
\end{figure}

In the ferromagnetic resonance linear polarized microwave magnetic field is used to study the resonance spectrum of magnetic material
\begin{align}
\textbf{H}_0=H_0\hat{\textbf{z}}+\hat{\textbf{x}} \delta H \cos \omega t,
\end{align}
$\delta H\ll H_0$. Mathematically, a linear polarized magnetic field can be written in a combination of circularly polarized magnetic field with opposite frequency 
\begin{align}
\hat{\textbf{x}} \cos \omega t= \frac{1}{2}\sum_{w=\pm\omega} (\hat{\textbf{x}}\cos w t+\hat{\textbf{y}}\sin w t).
\end{align}
Because of that, for mathematical simplicity, we can study the response of the magnetization dynamics of the following circularly polarized external magnetic field
\begin{align}
\textbf{H}_0=H_0\hat{\textbf{z}}+(\hat{\textbf{x}}\cos \omega t+\hat{\textbf{y}}\sin \omega t)\delta H .
\end{align}
The coupled magnetization dynamics of our ferrimagnet can then be linearized by setting $M_{j+}=M_{jx}+iM_{jy}$ and assuming $M_{j+}\ll M_{jz}$. The coupled dynamics can be written in the following linear equations.
%\begin{widetext}
\begin{equation}
\frac{\partial}{\partial t}
\left[
\begin{array}
[c]{c}%
M_{1+}\\
M_{2+}
\end{array}
\right]  = iW \left[
\begin{array}
[c]{c}%
M_{1+}\\
M_{2+}
\end{array}
\right] -ie^{i\omega t}\delta H \left[
\begin{array}
[c]{c}%
\left(1+i\alpha_1\right)\gamma_1M_1\\
\left(1+i\alpha_2\right)\gamma_2M_2
\end{array}
\right] \label{LinearLL}
\end{equation}
where $W=$
\begin{equation}
\left[
\begin{array}
[c]{cc}%
\left(1+i\alpha_1\right)\gamma_1\left(H_0+\lambda M_2\right) & -\left(1+i\alpha_1\right)\gamma_1\lambda M_{1}\\
-\left(1+i\alpha_2\right)\gamma_2\lambda M_{2} & \left(1+i\alpha_2\right)\gamma_2\left(H_0+\lambda M_1\right)
\end{array}
\right]
\end{equation}
%\end{widetext}
For $\lambda\gg H_0 $ one can show that the leading order in the eigen values of $W$ are 
\begin{align}
w_1=& \lambda\left(\left(1+i\alpha_2\right)\gamma_2M_1+\left(1+i\alpha_1\right)\gamma_1M_2\right) ,\\
w_2=& H_0 \frac{M_1+M_2}{\frac{M_1}{(1+i\alpha_1)\gamma_1}+\frac{M_2}{(1+i\alpha_2)\gamma_2}}.
\end{align}
In the limit $\lambda\gg H_0 $, the solution for $\delta H=0$ can be written as 
\begin{align}
{M}_{\mathrm{tot}+}\propto e^{i(1+i\alpha_\mathrm{eff})\gamma_\mathrm{eff} H_0 t},
\end{align}
as illustrated in Fig.~\ref{Fig.Ferri}. 
The eigenstate of $w_2$ determines the effective gyromagnetic ratio 
\begin{equation}
\gamma_\mathrm{eff}=\frac{\mathrm{Re} w_2}{H_0}
=
\frac{(M_1+M_2)\left(\frac{M_1/\gamma_1}{1+\alpha_1^2}+\frac{M_2/\gamma_2}{1+\alpha_2^2}\right)}{\left(\frac{M_1/\gamma_1}{1+\alpha_1^2}+\frac{M_2/\gamma_2}{1+\alpha_2^2}\right)^2+\left(\frac{\alpha_1M_1/\gamma_1}{1+\alpha_1^2}+\frac{\alpha_2M_2/\gamma_2}{1+\alpha_2^2}\right)^2} 
%\frac{(M_1+M_2)\left(\frac{1+\alpha_2^2}{\gamma_1}M_1+\frac{1+\alpha_1^2}{\gamma_2}M_2\right)}{\left(\frac{M_1}{\gamma_1}+\frac{M_2}{\gamma_2}\right)^2+\left(\frac{\alpha_2M_1}{\gamma_1}+\frac{\alpha_1M_2}{\gamma_2}\right)^2},  
\label{Eq.result}
\end{equation}
and the effective damping 
\begin{equation}
\alpha_\mathrm{eff}=\frac{\mathrm{Im} w_2}{\mathrm{Re} w_2} = 
\frac{\alpha_1\frac{M_1/\gamma_1}{1+\alpha_1^2}+\alpha_2\frac{M_2/\gamma_2}{1+\alpha_2^2}}{\frac{M_1/\gamma_1}{1+\alpha_1^2}+\frac{M_2/\gamma_2}{1+\alpha_2^2}}.
\end{equation}

$\alpha_\mathrm{eff}$ is closely related to the width of the ferromagnetic resonance (FMR) spectrum, which can be determined from
the rate of the loss of magnetic dissipation energy $\Delta F=-\textbf{H}_0\cdot \textbf{M}_\mathrm{tot}$.
\begin{align}
\frac{dF}{dt}= \frac{\alpha_\mathrm{eff}\gamma_\mathrm{eff}\omega^2 M_\mathrm{tot}\delta H^2}{\left(\omega-\mathrm{Re}w_2\right)+\left(\mathrm{Im}w_2\right)^2} 
\end{align}
The shape of the Lorentzian function indicates that the FMR width is proportional to the effective damping parameter
\begin{align}
\frac{\Delta \omega}{\omega_\mathrm{peak}}\sim\alpha_\mathrm{eff}.
\end{align}
In the limit of small $\alpha_1, \alpha_2\to 0$, we get the following well-known effective gyromagnetic ratio
\begin{align}
\lim_{\alpha\to0}\gamma_\mathrm{eff}= \frac{M_1+M_2}{{M_1}/{\gamma_1}+{M_2}/{\gamma_2}}. \label{Eq.gamma0}
\end{align}
On the other hand, in the limit of large $\alpha_2\gg \alpha_1\approx0$ we arrive at the Kittel gyromagnetic ratio for ferrimagnet with an overdamped $M_2$~\cite{Kittel1959,Kittel-VanVleck}
\begin{align}
\lim_{\alpha_2\to\infty}\gamma_\mathrm{eff}= \frac{M_1+M_2}{{M_1}/{\gamma_1}}.
\end{align}

\begin{figure}
\includegraphics[width=\columnwidth]{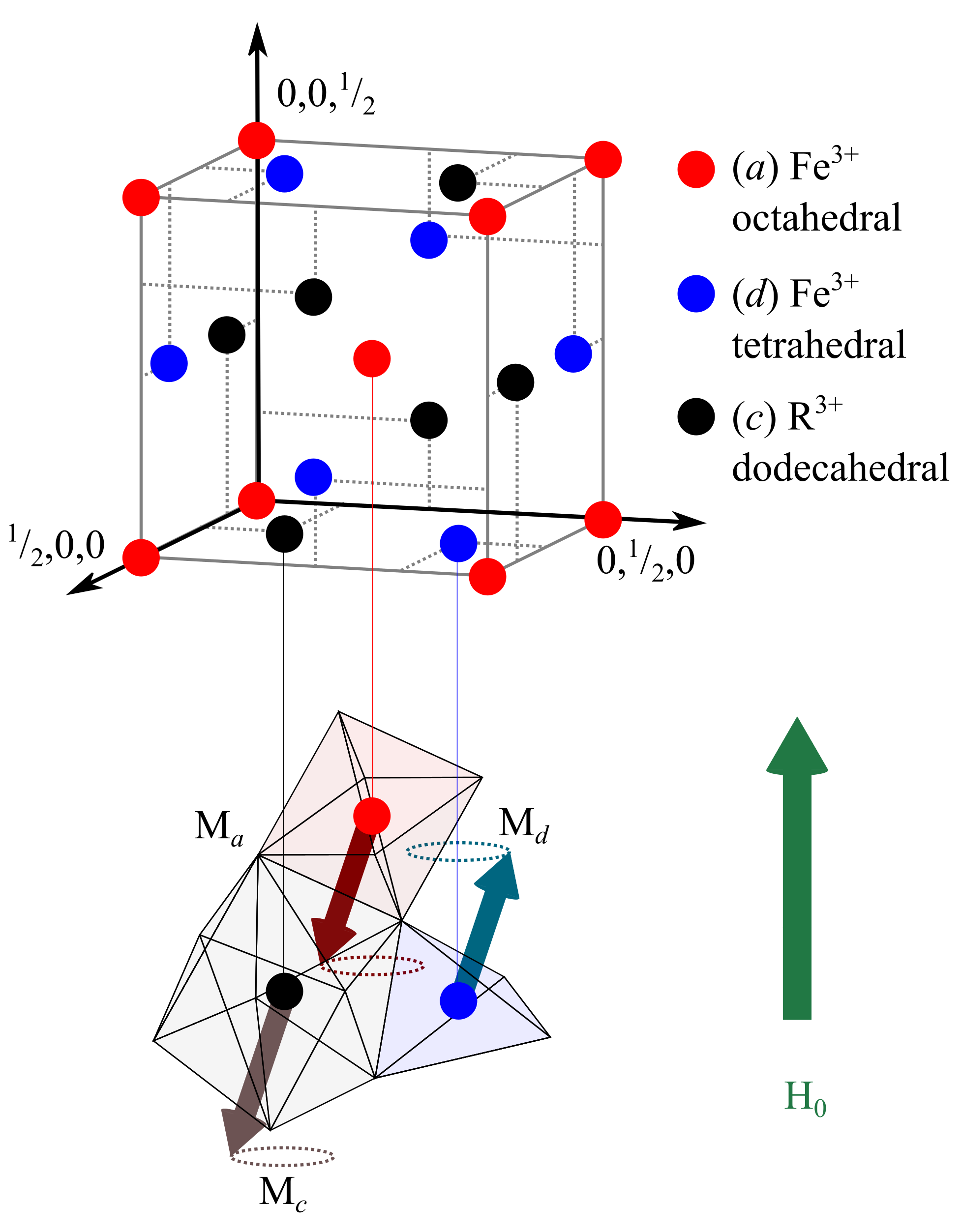}
\caption{One eighth of a unit cell of ferrimagnetic R$_3$Fe$_5$O$_{12}$. Lattice constant $a=12.4$ \AA. The garnet structure has metal cations (Fe$^{3+}$ and R$^{3+}$) and oxygen anions that form tetrahedral $(a)$, octahedral $(d)$ and dodecahedral $(c)$ sites \cite{GILLEO19801,Jain2013,Vesta}. R site is occupied by Y or rare earth elements and coupled antiferromagnetically with $a$-site. Fe ions occupy $a$ and $d$ sites. Because $a$ and $d$ are antiferromagnetically coupled, 4 out of 5 Fe$^{3+}$ in R$_3$Fe$_5$O$_{12}$ cancel each other.
\label{Fig.RIG}}
\end{figure}

\section{Results and Discussion}
\label{Sec:YGdIG}

From here on, we focus on substituted $\mathrm{Y}_3\mathrm{Fe}_5\mathrm{O}_{12}$. It has a garnet structure that consists of tetrahedron $(d)$, octahedron $(a)$ and dodecahedron $(c)$ of oxygen ions coordinated with metal cations. The magnetic moments in tetrahedral and octahedral sites rise from Fe$^{3+}$ ions \cite{GILLEO19801}. Because $a$ and $d$ sites are antiferromagnetically coupled, 4 out of 5 Fe occupying $a$ and $d$ sites cancel each other. Y in dodecahedral site can be substituted with rare earth elements and is coupled antiferromagnetically with $a$-site as seen in Fig.~\ref{Fig.RIG} \cite{GILLEO19801}.

Ref.~\cite{PhysRevB.87.104412} experimentally measures the gyromagnetic ratio of Y$_{3-x}$Gd$_{x}$Fe$_{5-y}$(Mn,Al)$_{y}$O$_{12}$ for variations of $x$ and $y$. Since Gd$^{3+}$ has non zero magnetization from half-filled 4$f$ orbital, substitution of Y creates magnetic moment at $c$ site. Mn$^{2+}$ can substitute Fe$^{3+}$ in $a$ site \cite{Geller1960}. Al dominantly substitute Fe$^{3+}$ in $d$ site when $y\leq 2$. For  $y= 6$\%, 90\% of Al$^{3+}$ substitutes $d-$site, this percentage reduces slowly as Al percentage increases \cite{GILLEO19801}. 
Main contribution of Mn and Al to the magnetization is the substitution of Fe in $a$ site \cite{GILLEO19801}. Fig.~\ref{Fig.M} illustrates that the magnetization of substituted Y$_3$Fe$_5$O$_{12}$ is dominated by Fe and Gd. 

\begin{figure}
\includegraphics[width=\columnwidth]{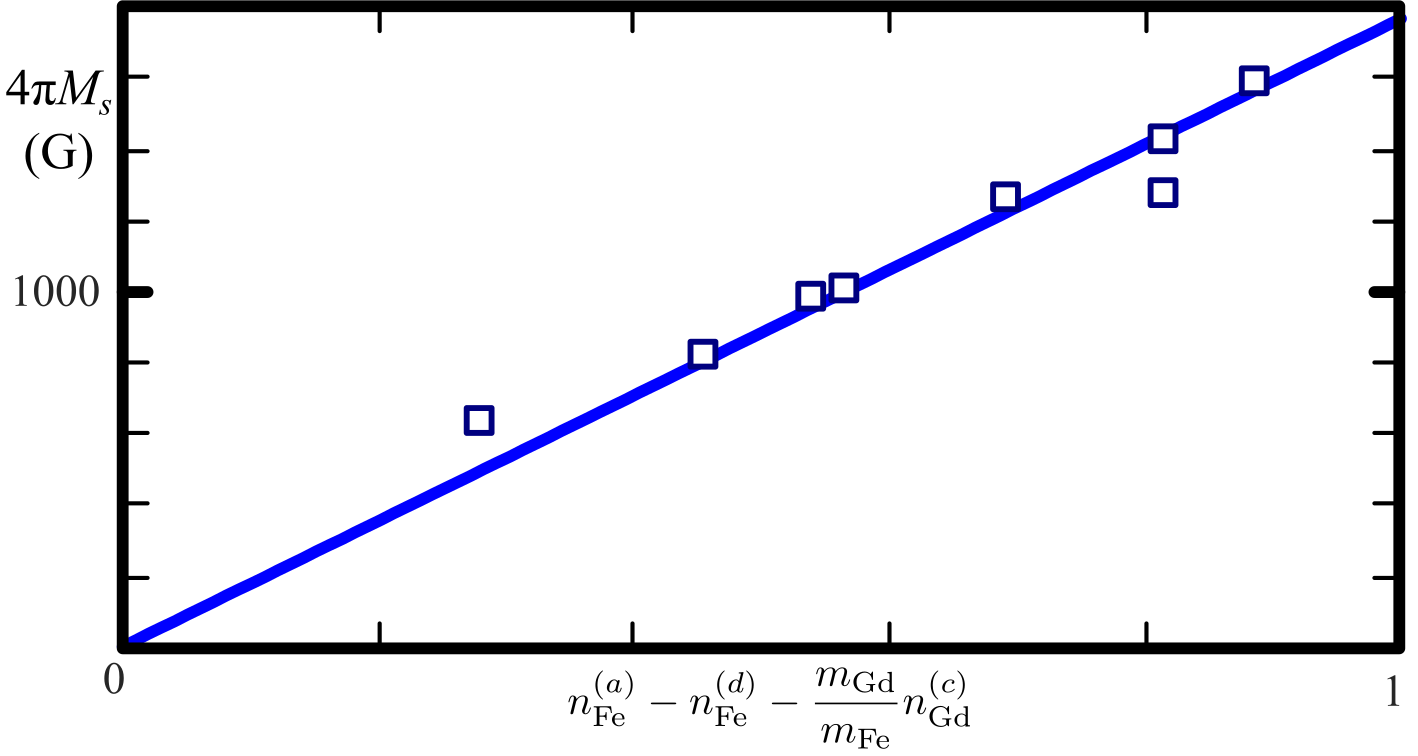}
\caption{ Magnetization of Y$_{3-x}$Gd$_{x}$Fe$_{5-y}$(Mn,Al)$_y$O$_{12}$ in Ref.~\cite{PhysRevB.87.104412} is dominated by Fe and Gd ions. Mn and Al indirectly contribute to the magnetization by substituting Fe \cite{GILLEO19801}. $M_s$ is saturation magnetization,  $m_{\rm X}$ is the magnetic moment of X ion. $n^{(j)}_X$ is the number of X ion at $(j)$-site. Here $m_{\rm Gd}=m_{\rm Fe}/3$.
\label{Fig.M}}
\end{figure}

Since magnetic moment at $d$ site cancels some of $a$ site, the magnetization of the Gd-substituted garnet arises from Fe$^{3+}$ of $a$ site and Gd$^{3+}$ of $c$ site. We can then set Fe$^{3+}$ of $a$ site to be the first magnetic lattice and Gd$^{3+}$ of $c$ site.
\begin{align}
M_1=&m_{\rm Fe}n_{\rm Fe} \label{Eq.MFe}, \\ 
M_2=&-m_{\rm Gd}n_{\rm Gd} \label{Eq.MGd},
\end{align}
On the other hand, $m_{\rm Gd}$ is the magnetic moment of Gd$^{3+}$. $m_{\rm Fe}$ is the magnetic moment of Gd$^{3+}$. Their ratio is determined by the paramagnetic response of Gd to the molecular field of Fe ion, according to Curie law \cite{NEEL1964344}
\begin{align}
\frac{m_{\rm Gd}}{m_{\rm Fe}}\propto\frac{1}{T}.
\end{align}
Since the compensation temperature and Curie temperature of Gd$_3$Fe$_5$O$_{12}$ is around 286 K \cite{PhysRev.137.A1034}
\begin{align}
M_{\rm Gd_3Fe_5O_{12}}=&m_{\rm Fe}-3m_{\rm Gd}=0,
\end{align}
one can estimate that $m_{\rm Gd}=m_{\rm Fe}/3$.

We can now describe the trend of gyromagnetic ratio using Eq.~\ref{Eq.result}. Fig.~\ref{Fig.YGdIG} illustrate the agreement of Eq.~\ref{Eq.result} with experimental data from Ref.~\cite{PhysRevB.87.104412}. The blue line is the gyromagnetic ratio of Y$_{3-x}$Gd$_{x}$ Fe$_{5-y}$(Mn,Al)$_{y}$O$_{12}$ bulk. From numerical fitting, one can find that {
\begin{align}
\alpha^{(0)}_{\rm Fe}=&0,\\
\alpha^{(0)}_{\rm Gd}=&0.36\pm 0.3,\\
\frac{\gamma_{\rm Fe}}{\gamma_{\rm Gd}}=& 1.43\pm 0.10.
\end{align}
}
The value of $\gamma_{\rm Gd}$ can be lower than $\gamma_{\rm Fe}$ because of the crystalline field \cite{PhysRev.181.478}.
For a bilayer of Y$_{3-x}$Gd$_{x}$ Fe$_{5-y}$(Mn,Al)$_{y}$ O$_{12}$ and Pt, we need to take into account the contribution of Re$G^{\uparrow\downarrow}$ according to Eq.~\ref{Eq.alphaS}. 

The spin mixing conductance at the interface Y$_{3-x}$Gd$_{x}$ Fe$_{5-y}$(Mn,Al)$_{y}$O$_{12}$ and Pt increases the magnetic damping of the ferrimagnet. { Since $
G^{\uparrow\downarrow}_{\rm YIG}=N_{\rm Fe}G_{\rm Fe}^{\uparrow\downarrow}$
, the damping enhancement of Fe lattice is
\begin{align}
\alpha_{\rm Fe}-\alpha^{(0)}_{\rm Fe}=\frac{N_{\rm Fe}G_{\rm Fe}^{\uparrow\downarrow}}{M_{\rm Fe}V}= \frac{G_{\rm YIG}^{\uparrow\downarrow}/A}{dM_{\rm YIG}}=0.95.
\end{align}
Here we used $d=1$ mm \cite{PhysRevB.87.104412} and lattice constant $\sim 12$ \AA. Using $G_j\propto S_j^2$ proportionality, the damping enhancement of Gd lattice can also be estimated
\begin{align}
\Delta\alpha_{\rm Gd}=&\frac{N_{\rm Gd}G^{\uparrow\downarrow}_{\rm Gd}}{M_{\rm Gd}V}= \frac{S^2_{\rm Gd}m_{\rm Fe}}{m_{\rm Gd}S^2_{\rm Fe}}\Delta \alpha_{\rm Fe}=0.22.
\end{align}
}
The change of damping parameter shifts the minimum value of the gyromagnetic ratio as seen in Fig.~\ref{Fig.YGdIG}. Since $\gamma$ that includes spin mixing contribution is extracted from $V$ in Ref.~\cite{PhysRevB.87.104412} using Eq.~\ref{Eq.SSE} (see Appendix \ref{Sec:App.bilayer}), the agreement with the experiment data confirms the proportionality of $V$ and $\gamma$.

\begin{figure}
\includegraphics[width=\columnwidth]{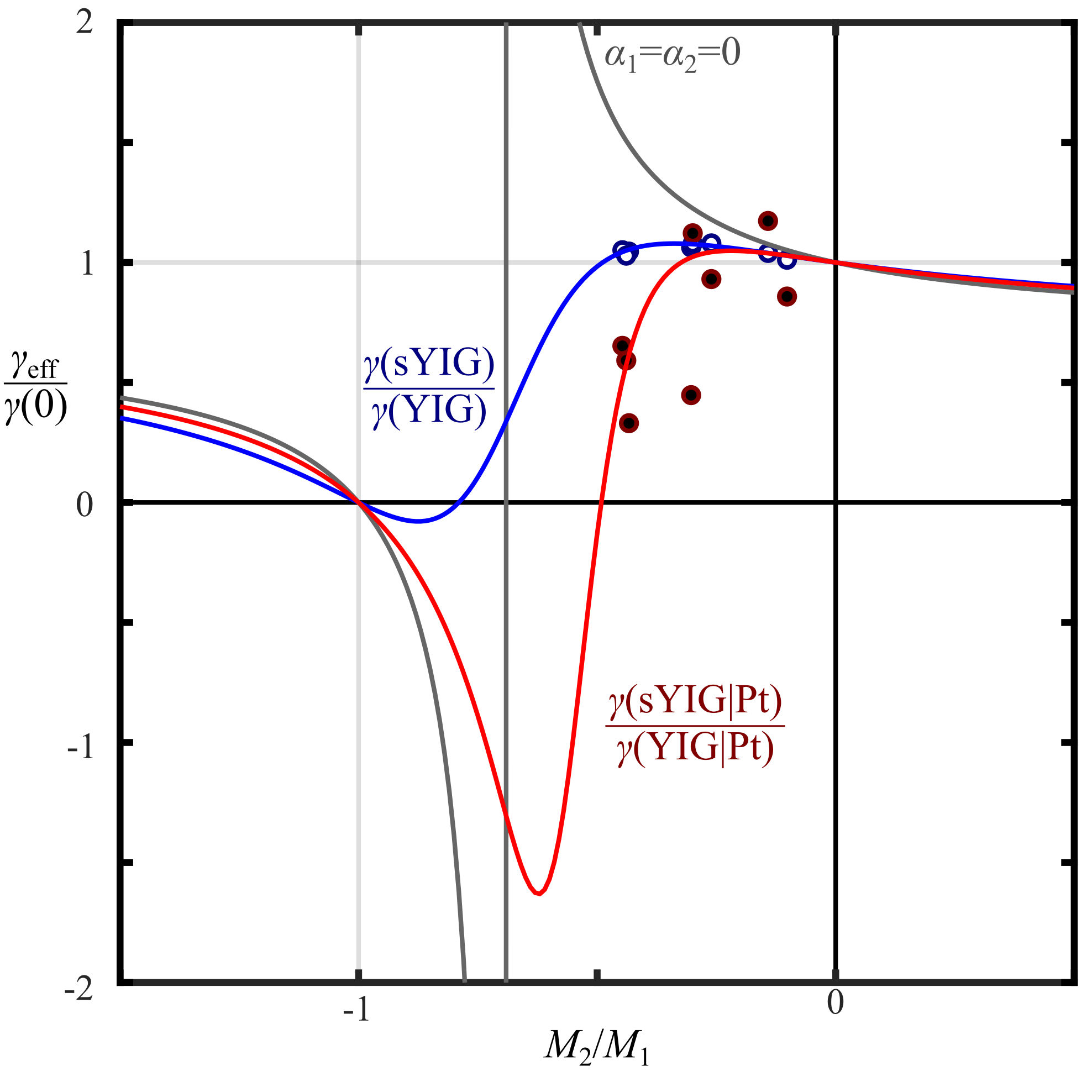}
\caption{Normalized gyromagnetic ratio of substituted YFe$_5$O$_{12}$ (sYIG) and sYIG|Pt as a function of the ratio of $M_1=m_{\rm Fe}n_{\rm Fe}$ and $M_2=-m_{\rm Gd}n_{\rm Gd}$. The values are normalized to the value of YFe$_5$O$_{12}$ $\sim$ 1.76$\times 10^{7}$ G$^{-1}$s$^{-1}$ \cite{doi:10.1063/1.5121013}.  Negative values of $M_2/M_1$ indicate that they are antiferromagnetically coupled. 
Blue line is the gyromagnetic ratio of sYIG with intrinsic damping $\alpha_{\rm Fe}=0$ and $\alpha_{\rm Gd}=0.36$. Red line is the gyromagnetic ratio of sYIG|Pt with increased {$\alpha_{\rm Fe}=0.95$ and $\alpha_{\rm Gd}=0.58$ } due to spin mixing at the interface. Theoretical values (blue and red lines) agree with the experimental values of sYIG (white circles) and sYIG|Pt (black circles) in Ref.~\cite{PhysRevB.87.104412} (see Table \ref{Table.data}).  The minimum value is shifted and the width is broadened. 
\label{Fig.YGdIG}}
\end{figure}

\begin{table}
\caption{Number of Gd$^{3+}$ and Fe$^{3+}$ that contributes to magnetization and gyromagnetic ratio of Y$_{3-x}$Gd$_{x}$Fe$_{5-y}$(Mn,Al)$_y$O$_{12}$ (sYIG) from Ref.~\cite{PhysRevB.87.104412}. The values of $\gamma_{\rm sYIG|Pt}$ are deduced from the spin Seebeck voltages (see Appendix \ref{Sec:App.bilayer}).}
\label{Table.data}
\begin{tabular}{cc|cc|cc}
\hline &	&	  & 	&  & \\[-1.5ex]
$x$ & $y$ & $n_{\rm Gd}$ & $n_{\rm Fe}$ & $\gamma_{\rm sYIG}$ & $\gamma_{\rm sYIG|Pt}$  \\[1ex]
 &  &  & & ($10^{7}$(Gs)$^{-1}$) & ($10^{7}$(Gs)$^{-1}$) \\[1ex]
\hline &	&	  & 	& \\[-1.5ex]
0.69 &	0.628  &   0.65  &    0.50 & 1.84 & 0.58 \\[1ex]
0.72 &	0.218  &   0.70  &    0.77 & 1.87 & 0.79 \\[1ex]
1.11 &	0.208  &   1.10  &    0.82 & 1.85 & 0.15 \\[1ex]
0.40 &	0.102  &   0.40  &    0.94 & 1.83 & 2.06 \\[1ex]
0.90 &	0.092  &   0.86  &    1.10 & 1.90 & 1.64 \\[1ex]
0.31 &	0.018  &   0.30  &    0.98 & 1.78 & 1.51 \\[1ex]
1.35 &	0.018  &   1.33  &    1.01 & 1.81 & 1.04 \\[1ex]
0.91 &	0.006  &   0.89  &    0.99 & 1.90 & 1.97 \\[1ex]
\hline
\end{tabular}
\end{table}

\section{Conclusion}

To summarize, we discuss the effect of spin mixing conductance on the effective gyromagnetic ratio of ferrimagnetic resonance of two magnetic lattices using Landau - Lifshitz equation. 
We apply the two lattices model to ferrimagnetic Y$_{3-x}$ Gd$_{x}$Fe$_{5-y}$(Mn,Al)$_{y}$O$_{12}$. The two lattices model can be used for the substituted Mn and Al substitution mainly replace Fe at $a$-site, and thus the magnetization only originated from Fe and Gd. We show that it can describe the effective gyromagnetic ratio of the substituted Y$_3$Fe$_5$O$_{12}$ with and without Pt interface. 

The interfacial spin mixing conductance influences the effective gyromagnetic ratio by increasing the damping parameter of Fe and Gd. Fig.~\ref{Fig.YGdIG} shows that the minima of gyromagnetic ratio of substituted Y$_{3}$Fe$_5$O$_{12}$ is further reduced due to spin mixing conductance of its interface with Pt. Far from the minima, the gyromagnetic ratio is weakly increased. As a comparison, the effect of small imaginary part of spin mixing conductance monotonically reduces the gyromagnetic ratio. 
Our result can be applied for Y$_3$Fe$_5$O$_{12}$ substituted by other rare earth elements which has various potential in spin-caloritonics and related areas.

%\section*{acknowledgments}
%The authors acknowledges the financial support from the Ministry of Research and Technology of the Republic of Indonesia through PDUPT Grant No. NKB-175/UN2.RST/HKP.05.00/2021.

\section*{DATA AVAILABILITY STATEMENTS}
The authors confirm that the data supporting the findings of this study are available within the article.

\appendix

{
\section{Relation of gyromagnetic ratio and spin Seebeck voltage in sYIG|Pt bilayer}

\label{Sec:App.bilayer}
Eq.~\ref{Eq.SSE} can be used for extracting the gyromagnetic ratio of sYIG|Pt bilayer from the spin Seebeck voltage 
\begin{align}
\frac{\gamma_{\rm sYIG|Pt}}{\gamma_{\rm YIG|Pt}}=\frac{\left(V M_s/G^{\uparrow\downarrow}\right)_{\rm sYIG|Pt}}{\left(V M_s/G^{\uparrow\downarrow}\right)_{\rm YIG|Pt}}.
\end{align}
From Eq.~\ref{Eq.GproptoS2}, one can arrive at 
\begin{equation}
G^{\uparrow\downarrow}_{\rm sYIG|Pt}\propto n_{\rm Fe}S^2_{\rm Fe}+n_{\rm Gd}S^2_{\rm Gd}.
\end{equation}
Because of that we can find the ratio 
\begin{align}
\frac{\gamma_{\rm sYIG|Pt}}{\gamma_{\rm YIG|Pt}}=&\frac{V_{\rm sYIG|Pt}}{V_{\rm YIG|Pt}} \frac{1+\frac{m_{\rm Gd}n_{\rm Gd}}{m_{\rm Fe}n_{\rm Fe}}}{1+\frac{S^2_{\rm Gd}}{S^2_{\rm Fe}}\frac{n_{\rm Gd}}{n_{\rm Fe}}}\notag\\
=& \frac{V_{\rm sYIG|Pt}}{V_{\rm YIG|Pt}}  \frac{1+\frac{m_{\rm Gd}n_{\rm Gd}}{m_{\rm Fe}n_{\rm Fe}}}{1+\frac{n_{\rm Gd}\left(m_{\rm Gd}/\gamma_{\rm Gd}\right)^2}{n_{\rm Fe}\left(m_{\rm Fe}/\gamma_{\rm Fe}\right)^2}},
\end{align}
which is useful for extracting $\gamma_{\rm sYIG|Pt}$ from raw spin Seebeck voltage data in Ref.~\cite{PhysRevB.87.104412} (see Table  \ref{Table.appendix}).
}

\begin{table}[hb]

\caption{Values of $n_{\rm Gd}$, $n_{\rm Fe}$, $V_{\rm sYIG|Pt}$, $\gamma_{\rm sYIG|Pt}$ extracted from Ref.~\cite{PhysRevB.87.104412} and the corresponding gyromagnetic ratio $\gamma_{\rm sYIG|Pt}$}
\label{Table.appendix}
\begin{tabular}{cccc}
\hline   & 	&  & \\[-1.5ex]
$n_{\rm Gd}$ & $n_{\rm Fe}$ & $V_{\rm sYIG|Pt}/V_{\rm YIG|Pt}$ & $\gamma_{\rm sYIG|Pt}$  ($10^{7}$(Gs)$^{-1}$) \\
\hline 
  0.65  &    0.50 & 0.50 & 0.58 \\
  0.70  &    0.77 & 0.60 & 0.79 \\
  1.10  &    0.82 & 1.00 & 0.15 \\
  0.40  &    0.94 & 1.35 & 2.06 \\
  0.86  &    1.10 & 1.20 & 1.64 \\
  0.30  &    0.98 & 0.95 & 1.51 \\
  1.33  &    1.01 & 0.90 & 1.04 \\
  0.89  &    0.99 & 1.50 & 1.97 \\
\hline
\end{tabular}
\end{table}

%\bibliography{ref}
%\bibliographystyle{apsrev4-1}


\end{small}
\end{document}